\documentclass[aps,pra,groupedaddress,amsmath,amssymb,twocolumn]{revtex4-1}

\usepackage{amsmath}
\usepackage[utf8]{inputenc}
\usepackage[T1]{fontenc}
\usepackage[english]{babel}
\usepackage[stable]{footmisc}
\usepackage[english]{hyperref}
\usepackage{multirow}
\usepackage{booktabs}
\usepackage{graphicx}
\usepackage{caption}
\usepackage{float}
\usepackage{units}
\usepackage{upgreek}
\usepackage{mathtools}
\usepackage{subfigure}
\usepackage{paralist}
\usepackage{xcolor}
\usepackage{amssymb}

\usepackage{todonotes}
\begin{document}

\title{Analysis of stochastic bifurcations with phase portraits}

\author{Marc Mendler}
 \email[]{marcm@fkp.tu-darmstadt.de} 
\author{Johannes Falk}
 \email[]{falk@fkp.tu-darmstadt.de} 
\author{Barbara Drossel}
 \email[]{drossel@fkp.tu-darmstadt.de}

\affiliation{Institut für Festkörperphysik, Technische Universität Darmstadt, Hochschulstr. 6, 64289 Darmstadt, Germany}

\date{\today}

\begin{abstract}
We propose a method to obtain phase portraits for stochastic systems. Starting from the Fokker-Planck equation, we separate the dynamics into a convective and a diffusive part. We show that stable and unstable fixed points of the convective field correspond to maxima and minima of the stationary probability distribution if the probability current vanishes at these points. Stochastic phase portraits, which are vector plots of the convective field, therefore indicate the extrema of the stationary distribution and can be used to identify stochastic bifurcations that change the number and stability of these extrema. We show that limit cycles in stochastic phase portraits can indicate ridges of the probability distribution, and we identify a novel type of stochastic bifurcations, where the probability maximum moves to the edge of the system through a gap between the two nullclines of the convective field.

% \tdJF{
% In Anbetracht der dir geschickten Paper habe ich versucht den Abstract etwas milder zu formulieren:

% Phase portraits are a helpful tool in analysing dynamical systems. Nevertheless, for stochastic system noise and diffusion come into play and can drastically alter the behaviour of a dynamical system hence conventional phase portraits lose their informational value.
% We present a powerful method to draw phase diagrams of stochastic systems. Based on known properties of the stochastic system, we separate its dynamics into a convective and a diffusive part. Using known features of the Fokker-Planck equation we are then able to draw stochastic phase portraits that indicate the most probable directions in phase space. This method enables us to identify qualitatively different regimes of stochastic dynamics, as well as to classify new types of stochastic bifurcations.

% }

\end{abstract}

% insert suggested PACS numbers in braces on next linex

\maketitle %\maketitle must follow title, authors, abstract and \pacs

% \tdBD{folgende Englisch-Fehler kommen mehrfach vor: \\
% (i)Komma vor that (da kommt kein Komma hin. allgemein gilt bei Relativsaetzen: wenn der Hauptsatz auch ohne den Relativsatz seine Aussage macht, kann man Kommata setzen, aber dann verwendet man eher "which" als "that". Wenn der Hauptsatz ohne den Relativsatz nicht verstaendlich ist, dann setzt man keine Kommata). Wenn "that" "dass" bedeutet, dann kommt sowieso kein Komma hin, weil der Satz ohne den Nebensatz keinen Sinn macht.\\
% (ii) it's: das ist die Abkuerzung fuer "it is", nicht der Genitiv von "it", den du meinst. Den schreibt man "its".\\
% (iii) Bindestriche bei zusammengesetzen Ausdruecken: wenn ein Wort aus zwei Teilen besteht, kommt kein Bindestrich rein (also: "Schloegl model"). Bei drei Teilen kommt ein Bindestrich zwischen die ersten beiden, also "Fokker-Planck equation; Kramers-Moyal expansion".}

% \tdJF{
% (iv)Nach Adverbien wie: in fact, therefore, nevertheless, moreover, furthermore, still, instead wird ein Komma gesetzt.
% }

\section{Introduction}
%\subsection*{Dynamical Systems}
Networks of interacting species play an important role in many disciplines like for example chemistry, ecology, or systems biology \cite{sandefur_network_2013}. A powerful modeling approach for such networks are ODE (ordinary differential equation) models, which are often written in the form of a dynamical system,
\begin{equation}
\frac{\text d\vec{x}}{\text d t} = \vec{f}(\vec{x})\, .
\end{equation}
The function $\vec{f}(\vec{x})$ is called the deterministic drift and has the mathematical structure of a vector field \cite{hirsch2012differential}.
\smallskip

In particular in one- and two-dimensional systems, the general dynamical behavior can be directly deduced from vector plots of $\vec{f}(\vec{x})$ \cite{Nolte_2010}. Based on these, the qualitatively different types of trajectories and the invariant manifolds (fixed points, limit cycles) can be graphically represented in the form of so-called phase portraits \cite{perko2013differential}.

When control parameters, such as reaction rates, are changed, dynamical systems can undergo bifurcations, during which fixed points or limit cycles are destroyed or generated, or they change their stability. For instance, when the rate of predation in a predator-prey system is decreased, eventually a point will be reached where the predator can no longer find enough food and becomes extinct. Local bifurcations can be calculated from $\vec{f}(\vec{x})$ by determining the fixed points and their stability in dependence of the parameters. Global bifurcations, which destroy or generate limit cycles of a finite size, are much more difficult to deduce from $\vec{f}(\vec{x})$. In order to obtain a general first overview over the different types of bifurcations that occur in the system, it is useful to plot phase portraits for different values of the control parameters and to compare them to each other. 

%\subsection*{Stochastic Systems}
The above deterministic and continuous description is an idealization that is often too simple \cite{kampen, gardiner, gillespie_buch}. ODE models are based on mean reaction rates and are a good approximation when concentrations are large and systems are well mixed. They have been applied successfully to many different biological and chemical networks \cite{lotka1926elements, rosenzweig, brusselator}. However, species concentrations cannot be treated as continuous quantities when numbers become small, as there are no half molecules, proteins, or animals. In this case, one should use a discrete state space. Furthermore, the randomness of reactions can lead to considerable deviations from the deterministic dynamics \cite{bokes_gene_2016, tsimring_noise_2014, raj_stochastic_2008}. For instance, it can lead to accidental extinctions of species, to the appearance of quasi-cycles in predator-prey models \cite{mcKane_oscillations}, or to switching between two types of behavior, as in the  foraging efforts of ant colonies \cite{mcKane_ameisen}. Additionally, the same deterministic equation can arise from different microscopic models, either by using a reaction that produces one molecule often, or a reaction that produces several molecules at once, but less often. Such a bursty reaction occurs for example during translation in protein synthesis \cite{raj_stochastic_2008,friedman_linking_2006, lin2016bursting}. Different burst sizes lead to a different intrinsic noise of the system: Not only is the variance of the noise distribution proportional to the burst size, but the noise becomes asymmetric when only production but not the destruction of molecules is bursty. Such an asymmetry can shift the mean of the stationary distribution with respect to the value expected from the ODE model. \cite{grima, kampen}

In order to assess the relevance of stochasticity, it is useful to express the variables in terms of the system size or reactor volume $N$, which is a measure for the total number of reactants in the system. It is well known that the fluctuations in a stochastic system scale only with $\sqrt{N}$, whereas the mean values scale with $N$. This means that the signal-to-noise ratio decreases with $N$. In the limit $N \rightarrow \infty$, the ODE description is recovered. \cite{kampen}

There exist several modeling techniques for such stochastic systems, which describe the system on different levels of resolution. The most important of these techniques is the Master equation, which models the time evolution of the numbers of the different molecules as a Markov process with a discrete state space and continuous time \cite{gardiner}. As the time evolution depends only on the present state and not on the past, memory effects are ignored. Furthermore, a well-mixed system is assumed since no spatial effects are considered. 
Despite its simplicity, the Master equation cannot usually be solved analytically. For that reason, various approximations are used, which simplify the calculations but at the same time preserve important features of the stochastic description. 

One of these approximations is the Fokker-Planck equation, which is obtained from the Master equation by keeping the lowest-order terms of the so-called Kramers-Moyal expansion \cite{risken_fokker-planck_1984}. This expansion uses a coarse-grained, continuous state space and is essentially a Taylor expansion to second order in the changes of these continuous state variables. The Fokker-Planck equation becomes more accurate when more molecules are in the system. It works however still astonishingly well for systems with only a few tenths of molecules, even when ODE descriptions are already very bad \cite{grima2011accurate}. The Fokker-Planck equation will also be the starting point for our investigation. A competing approach is the system-size expansion \cite{kampen}, which uses the inverse system size as an expansion parameter. It is often considered to be more systematic than the Fokker-Planck equation, but nevertheless frequently leads to results which are worse, at least when carried out only to the first two terms \cite{grima2011accurate}.  

Besides those generic approaches, a vast amount of more specific techniques have been developed, which allow for all kinds of analyses of stochastic systems, such as effective deterministic descriptions \cite{grima_effective_2010}, most probable dynamics \cite{most_probable, moon_interpretation_2014}, stability analysis \cite{scott}, mean switching times of multistable systems \cite{gardiner}, stochastic resonant cycles \cite{mckane2005predator}, or piecewise deterministic Markov processes \cite{davis1984piecewise, lin2016bursting}. 

%\subsection*{Context of this paper}

So far, these techniques were rarely used to investigate the effect of stochasticity on the bifurcations of reaction networks. When the number of molecules is large and the dynamics resemble that of the deterministic ODE system, attractors of the dynamical system give rise to sharp maxima of the stationary probability density of the stochastic system. For this reason, bifurcations of stochastic systems are often understood as qualitative changes in the topology of the maxima and ridges of the stationary distribution. Those maxima can merge or split, such that a multistable distribution becomes monostable or vice versa.  This class of bifurcations is called phenomenological bifurcations, or p-bifurcations  \cite{arnold2013random}. We have investigated a stochastic bifurcation of this type in an analysis \cite{johannes_schloegl} of the Schloegl model \cite{schloegl_original} where we showed that burst noise shifts the parameter values of the saddle-node bifurcation. Burst noise can thus lead to bistability in a parameter region where the deterministic model or the model with conventional noise is monostable, and vice versa.
Similar effects were found in other types of systems, such as  in noisy bistable fractional-order systems \cite{yang_stochastic_2016}, or for the Duffing-Van der Pol oscillator \cite{yang_stochastic_2017,zakharova_stochastic_2010}.

Besides p-bifurcations, Arnold \cite{arnold2013random} defines dynamical bifurcations (d-bifurcations). Dynamical bifurcations correspond to a separation of dynamics into state space regions between which no transitions are possible. This definition is chosen in analogy to bifurcations in ODE systems. For this reason, several subsequent publications \cite{hopf_hoppe, Schenk1996} concentrate mainly on d-bifurcations. However, these bifurcations can occur only under the condition that the intrinsic noise of the system vanishes at the boundary between the different dynamical regions.  This cannot happen for the kind of chemical reaction networks that we are interested in.
% Unterschied additive/multiplicative noise: d-bifurcations entstehen nur bei multiplicative noise: http://www.alsharawi.info/Forms/Talks/MartinRasmussenICDEA2013A.pdf
Therefore, we focus on p-bifurcations, which are very common in stochastic reaction networks. Since the stationary distribution of the probability density contains no information about the dynamics of the system, our study will not be confined to the calculation of stationary densities. Instead, we want to suggest a stochastic analogon of phase portraits, which convey a better understanding of the dynamics of the system and give an immediate insight into the possible bifurcations. A similar idea has been pursued in \cite{most_probable}, where the "most probable" trajectories are calculated, starting from an initial delta distribution. In contrast, our method gives information on the  probability flows everywhere in state space starting from a constant distribution and is therefore not limited to unimodal distributions.  This is, of course, crucial to the investigation of stochastic bifurcations. We will discuss the connection between these two formalisms further in the final section.

%\subsection*{Structure of this paper}
This article is structured as follows:
Section \ref{sec:methods} summarizes the methods that will be used, introduces stochastic phase portraits and shows that extrema of the stationary probability distribution coincide with fixed points of the convective field when the probability current vanishes.
%, beginning with a short definition of chemical reaction networks in section \ref{sec:crn}, followed by some explanations about the multidimensional Fokker-Planck-Equation (section \ref{sec:multi_FPE}) that is the starting point for our analysis. In Section \ref{sec:favorable_states}, we will introduce favorable and unfavorable states of the stochastic system, which are essentially extrema of the stationary Fokker-Planck-Equation.
%In this context, we will also introduce the convective field $\vec{\alpha}(\vec{x})$ that is the basis of creating stochastic phase portraits (SPPs), which will be introduced in section \ref{sec:spps}.

The second part of the paper gives several examples for the application of these phase portraits, demonstrating the power of the method. Thus, the bifurcation diagram of a model for foraging ant colonies  \cite{mcKane_ameisen} will be analyzed without need to solve the Fokker-Planck equation, and the relation between limit cycles of the convective field and ridges of the probability distribution will  be explored using the Rosenzweig-MacArthur predator-prey model. 
%In section \ref{sec:ants}, we start with a well-understood example about a foraging ant colony \cite{mcKane_ameisen}, which can be studied in a very straightforward way with our method. Section \ref{sec:rosenzweig} contains a more complex model that shows the full potential of the stochastic phase portraits. 
As part of this investigation, we will encounter a special type of bifurcation that can only emerge in stochastic systems and has no deterministic counterpart. Finally, we will summarize and discuss our results in section \ref{sec:discussion}.

\section{Methods}
\label{sec:methods}

\subsection{Chemical reaction networks}
\label{sec:crn}

We consider reaction networks for species  $X_i$ that undergo a set of reactions
\begin{align}
\begin{aligned}
\label{eq:reaction_system}
\sigma_{11} X_1 + \sigma_{21} X_2  +\dotso + \sigma_{k1} X_k &\overset{\mu_1}{\longrightarrow} \rho_{11} X_1 + \dotso + \rho_{k1} X_k\\
&\vdots\\
\sigma_{1m} X_1 + \sigma_{2m} X_2 + \dotso + \sigma_{km} X_k &\overset{\mu_m}{\longrightarrow} \rho_{1m} X_1 + \dotso + \rho_{km} X_k\\
\end{aligned}
\end{align}
with so-called stoichiometric constants $\sigma_{ij}$ and $\rho_{ij}$ and reaction rates $\mu_j$.

To compactify the notation, one defines \cite{kampen} the stoichiometric matrix
\begin{equation}
	S_{ij} = \rho_{ij} - \sigma_{ij}
\end{equation}
 and the propensity vector
\begin{equation}
	\nu_j(\vec{n}) = \mu_j \prod_{z=1}^k  N^{-\sigma_{zj}}  \frac{n_z!}{(n_z-\sigma_{zj})!}\, ,
\end{equation}
which equals the reaction rate $\mu_j$ per molecule(s) times the probability of the involved molecules to meet. In this notation $n_i$ is the molecule number of species $X_i$ and $N$ is the system size or reactor volume, which can be interpreted as an inverse measure of the intrinsic noise of the system.

With these definitions, the set of reactions (\ref{eq:reaction_system}) can be written in form of the so-called (chemical) master equation,
\begin{equation}
\label{eq:cme}
\frac{d p(\vec{n}, t)}{dt} = N \sum_{j = 1}^m \left( \prod_{i = 1}^{k} \mathbb{E}_i^{-S_{ij}} - 1 \right) \nu_j (\vec{n}, N) p (\vec{n}, t)
\end{equation}
with the step operator $\mathbb{E}_i$, which is defined via
\begin{equation}
\mathbb{E}^{\alpha}_i \phantom{\cdot} f(n_1, n_2, ..., n_i, ..., n_k) = f(n_1, n_2, ..., n_i + \alpha, ..., n_k)
\end{equation}
for an arbitrary function $f$ and an integer $\alpha$.

\subsection{The multidimensional Fokker-Planck equation}
\label{sec:multi_FPE}
The master equation yields a time-continuous, discrete-state Markov model of the system. As mentioned above, the analysis of the dynamics becomes simpler when using a continuous state space. Therefore we introduce the species concentrations $x_i = n_i / N$ and base our analysis on the multidimensional Fokker-Planck equation 
(FPE) \cite{gardiner}
\begin{align}
\begin{aligned}
\frac{\partial p(\vec{x},t)}{\partial t} &= - \sum_i \frac{\partial}{\partial x_i}\left[f_i(\vec{x})p(\vec{x},t)\right]\\
&+\frac{1}{2 N}\sum_{ij}\frac{\partial^2}{\partial x_i \partial x_j}\left[D_{ij}(\vec{x})p(\vec{x},t)\right]\, .
\label{eq:fpe}
\end{aligned}
\end{align}
Here, we have introduced the deterministic drift 
\begin{equation}
\vec{f}(\vec{x}) = \mathbf{S} \cdot \nu(\vec{x} \cdot N)
\end{equation}
% \tdBD{hier stimmt was nicht. In der Notation mit den Konzentrationen muss $\vec f$ doch von den Reaktionsraten $\mu$ und nicht von den Propensitaeten $\nu$ abhaengen. Ich denke, dass hier die Ursache fuer meine spaeteren Probleme in Teil III liegt - siehe die Kommentare in IIIB}
and the diffusion matrix
\begin{equation}
\mathbf{D}(\vec{x}) = \mathbf{S} \cdot \text{diag}(\nu) \cdot \mathbf{S}^T \;. 
\end{equation}
From the definition of (positive) definiteness ${\langle x, \mathbf{A}\cdot x \rangle > 0}$, it follows that 
$\mathbf{D}$ is positive definite. 

The drift term alone determines the ODE description of the system, which becomes exact in the limit of infinitely many molecules,
\begin{equation}
\dot{\vec{x}} = \vec{f}(\vec{x})\, .
\label{eq:dynamical_system}
\end{equation}

The FPE has the form of a continuity equation, $\frac{\partial p}{\partial t} = - \vec{\nabla} \cdot \vec{j}$ with the probability current 
\begin{equation}
j_i = \underbrace{\left(f_i(\vec{x}) - \frac{1}{2 N} \sum\limits_{k}  \frac{\partial D_{ik}}{\partial x_k}\right)}_{\alpha_i(\vec{x})} p(\vec{x},t) - \frac{1}{2 N}\sum\limits_k D_{ik}\frac{\partial p}{\partial x_k}\, .
\end{equation}
Defining 
\begin{equation}
\vec{\alpha}(\vec{x}) = \vec{f}(\vec{x}) - \frac{1}{2 N} \sum\limits_{ik}  \frac{\partial D_{ik}}{\partial x_k} \vec{\epsilon_i} = 0
\label{eq:alpha}
\end{equation}
with $\vec{\epsilon}_i = \vec{x}_i/|x_i|$ being the unity vector in direction $x_i$, the current can be written as
\begin{equation}
\vec{j} = \vec{\alpha}(\vec{x}) p(\vec{x},t) - \frac{1}{2 N}\mathbf{D}\cdot\vec{\nabla}p(\vec{x},t)\, .
\label{eq:j}
\end{equation}

The first term $\vec\alpha_i(\vec{x}) p(\vec{x},t)$ is usually called the convection term, the second one the diffusion term \cite{Schwabl2006}. To avoid confusion with the drift $\vec{f}$ and diffusion $\mathbf{D}$ terms in the FPE, we will refer to these two contributions to the probability current as the convective and diffusive current. 

The expression for the convective current $\vec{j}_c = p \vec{\alpha}$ has the same form as that for the electric current, $\vec{j} = \sigma \vec{E}$, and we therefore call $\vec\alpha$ the convective field. This field depends only on $\vec f$ and $\mathbf{D}$, but not on the probability density $p(\vec x,t)$. Its effect is to pump probability towards the preferred states of the system, just like an electric field pushes positive charges towards its sinks. In absence of the diffusive current, the convective current would converge to the attractors of the vector field $\vec \alpha$, producing a stationary probability distribution that is a (properly scaled) delta distribution on the attractors. 
The diffusive current, on the other hand, is directed toward decreasing $p$ and thus tends to flatten steep slopes, ironing out sharp initial peaks and giving rise to a finite width of the stationary distribution. When a constant probability density is chosen as initial distribution, the diffusive current vanishes, and the convective field gives the direction of the current during the first infinitesimal time interval. 

\begin{figure}
    \includegraphics[width=0.95\linewidth]{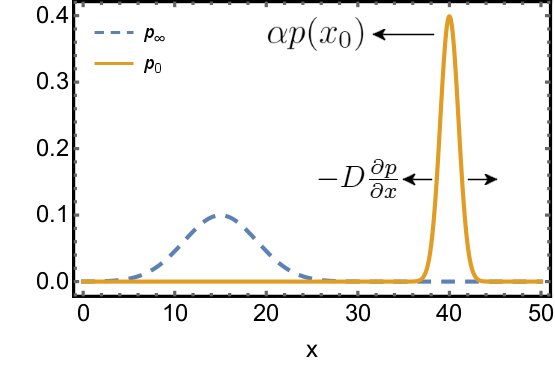}    
\caption{Illustration of the two contributions to the current in (\ref{eq:j}). The sharp initial distribution $p_0$  (solid orange line) is driven by the convective current $\alpha p(x_0)$ towards the stationary stable state $p_\infty$ (dashed blue line), whereas the diffusive current $-D\frac{\partial p}{\partial x}$ broadens the peak.
}
\label{fig:j_meaning}
\end{figure}

The stationary distribution is given by the condition $\frac{\partial p(\vec{x},t)}{\partial t}=0$, which means $\vec{\nabla} \cdot \vec{j}=0$. In a one-dimensional closed system, where $\vec j$ must vanish at the boundaries, this means that $\vec j=0$ in the stationary state. The convective and diffusive currents cancel each other. 
Fig.~\ref{fig:j_meaning} illustrates the effect of the two fields on a narrow initial distribution for a one-dimensional system. The initial peak moves towards the stationary solution and becomes broader until in the stationary state the two currents cancel each other.  
In more than one dimension, stationary solutions with $\vec j \neq 0$ are also possible, as the condition  $\vec{\nabla} \cdot \vec{j}=0$ can be satisfied with nonzero currents.  Such states are often called non-equilibrium steady states. The condition $\vec{\nabla} \cdot \vec{j}=0$ implies for these states that $\vec j = \vec \nabla \times \vec A$ with a vector field $\vec A(\vec x)$, i.e., $\vec{j}$ is solenoidal.

% In the following we will use the abbreviations $\vec{A}$ and $\vec{D}$ for theses quantities and write
% \begin{equation}
% \vec{j} =\vec{A} + \vec{D}
% \end{equation}
% \smallskip

%\subsection{Properties of the convective field}
%\label{sec:convection}

\subsection{Favorable and unfavorable states}
\label{sec:favorable_states}

We define \textit{favorable states} as maxima of the stationary probability density at which the probability current $\vec j$ vanishes. At these maxima, we have $\frac{\partial p}{\partial x_k} = 0 \; \forall k$ in (\ref{eq:j}). The condition $\vec j=0$ applied to (\ref{eq:j}) gives $\vec\alpha=0$. This means that favorable states are fixed points of $\vec\alpha$.

The calculation of these favorable states can be done without explicitly solving the FPE. In fact, this calculation is mathematically almost identical to the calculation of the fixed points in the deterministic description, which are obtained from $\vec{f}(\vec{x}) = 0$. The only difference is that $\vec f$ is replaced with $\vec \alpha$ in the stochastic description. 

In analogy to unstable fixed points of deterministic systems, we also consider minima of the stationary distribution $ p_{\infty}(\vec x)$ at which the probability current vanishes. These are also fixed points of $\vec \alpha$, and we call them  \textit{unfavorable states} of the system. 

Favorable and unfavorable states are associated with a different stability of the fixed point of $\vec\alpha$: 
Setting \eqref{eq:j} equal to zero and differentiating with respect to a component $x_m$, we obtain
\begin{equation}
\frac{\partial \alpha_i}{\partial x_m}p + \alpha_i \frac{\partial p}{\partial x_m} = \frac{1}{2 N}\sum\limits_k\left(\frac{\partial D_{ik}}{\partial x_m}\frac{\partial p}{\partial x_k} + D_{ik} \frac{\partial^2p}{\partial x_k\partial x_m}\right)\, .
\end{equation}
Using $\frac{\partial p}{\partial x_i}=0$ gives
\begin{equation}
\frac{\partial \alpha_i}{\partial x_m}p  = \frac{1}{2 N}\sum\limits_jD_{ik} \frac{\partial^2p}{\partial x_k\partial x_m}\, .
\end{equation}
This can be rewritten as a matrix equation
\begin{equation}
\mathbf{h} = 2 N p \; \boldsymbol{D}^{-1} \cdot \boldsymbol{J_s}
\end{equation}
with $(\mathbf{J}_s)_{ik} = \frac{\partial \alpha_i}{\partial x_k}$ and $h_ {ik}=\frac{\partial^2 p}{\partial x_i x_k}$.
Since $\mathbf{D}$ is positive definite, its inverse is also positive definite. The multiplication with a positive definite matrix does not change the definiteness of a matrix. This means that the Hesse matrix  $\mathbf{h}$ is positive definite when $\mathbf{J_s}$ is positive definite, and negative definite when $\mathbf{J_s}$ is negative definite. It follows that favorable (unfavorable) states are stable (unstable) fixed points of $\vec\alpha$. The classification of these states can therefore be done with the same type of calculation as in the deterministic model, only with $\vec f$ being replaced with $\vec \alpha$. 

\smallskip

\begin{figure}
    \subfigure[$\vec{\alpha}(\vec{x})$]{\includegraphics[width=0.35\linewidth]{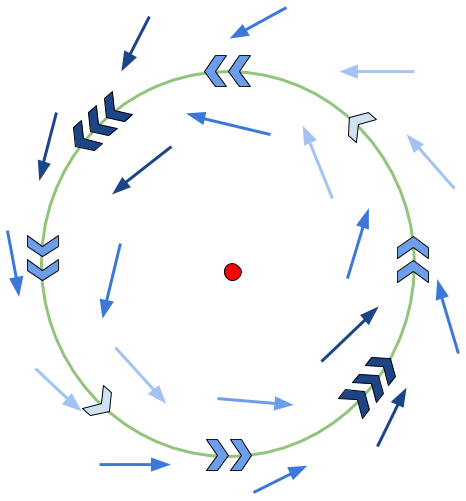}}
    \subfigure[$p(x,y)$]{\includegraphics[width=0.54\linewidth]{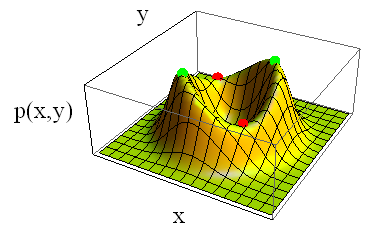}}
\caption{Schematic vue of the convective field $\vec{\alpha}(\vec{x})$ and stationary probability distribution for a system with a non-uniform limit cycle. Regions on the limit cycle with fast dynamics (implied by more and darker arrows in (a) lead to saddle points in the probability distribution  (red dots in (b)). Regions with slower dynamics (brighter and fewer arrows in (a)) lead to the formation of probability maxima (green dots in (b)). These types of maxima are \textit{slow transient states}.
}
\label{fig:methods:transient_maxima}
\end{figure}

There are also maxima $\vec{x}^*$ of the stationary probability density at which $\vec{\nabla}p_{\infty}(\vec{x}^*) = 0$ but $\vec{j}(\vec{x}^*) \ne 0$. These points are not fixed points of $\vec{\alpha}$. An example is given in fig. \ref{fig:methods:transient_maxima}. If the deterministic system has a stable limit cycle with an unstable spiral in the center, the stochastic system can be expected to have an  unfavorable state sitting in a crater surrounded by a ridge. Typically, the current along this ridge will not be constant, being faster at saddle points and slower at local maxima of the stationary probability density. These maxima cannot be   classified as favorable states of the system, as trajectories are not attracted to them but pass through them. They are rather \textit{slow transient states}. 

% \tdMM{Vermutung: slow transient states können nur auftreten, wenn rot$(\alpha) \ne 0$}

The reverse situation, that  $\vec{\alpha}(\vec{x_0}) = 0$ but that $\vec{x_0}$ is no extremum of $p_{\infty}$ can also be imagined. The current density at such a point must be different from zero. Although we never encountered such a "false positive" in the models we investigated, we could so far not show under which conditions they may occur or whether they can be ruled out completely. We will therefore always compare the calculated fixed points of $\vec{\alpha}$  with simulation results of the stochastic system, to verify that such a fixed point  indeed corresponds to a favorable state of the system.

\smallskip

\smallskip

\subsection{Stochastic phase portraits}
\label{sec:spps}

We will explore further below the usefulness of what we want to call stochastic phase portraits (SPPs), which show the qualitatively different types of trajectories defined by the vector field $\vec\alpha(\vec x)$. This means that we consider the dynamical system
\begin{equation}
\frac{\text d \vec{x}}{\text d t} = \vec{\alpha}(\vec{x})\, ,
\label{eq:pseudo_dynamical_system}
\end{equation}
for which we draw (conventional) phase portraits. As shown above,  the stable fixed points of \eqref{eq:pseudo_dynamical_system} coincide with the favorable states of the underlying stochastic system when the probability current vanishes at these points. Furthermore, we will use the trajectories calculated by solving \eqref{eq:pseudo_dynamical_system} numerically as an indicator of the probability flow in the system. This will be especially useful to determine the occurrence and position of stochastic limit cycles.

As these trajectories include only the convection flow of the probability current and are completely ignorant of the initial conditions of the stochastic system, they can, however, provide no information about the time evolution of a stochastic system with a specific initial distribution. 
In particular, one must not confuse these trajectories with sample paths of the stochastic system, nor with the evolution of its mean or of its maximum. The latter was investigated in a study of the most probable dynamics in \cite{most_probable}.

\section{Results}
\subsection{Example: Foraging colony}
\label{sec:ants}
In \cite{mcKane_ameisen} the authors propose a model for the emergence of bistability in the foraging behavior of an ant colony, which we want to use as a first application of the SPPs. The model introduces two ant "species" $X_1$ and $X_2$ that feed on two different food sources. There are two types of "reactions". The first one 
\begin{align}
\begin{aligned}
X_1 + X_2 &\overset{r}{\longrightarrow} 2 X_1\\
X_1 + X_2 &\overset{r}{\longrightarrow} 2 X_2
\end{aligned}
\end{align}
corresponds to the recruiting of an ant of one species by the other species, so that it changes its feeding source. The other type of reaction
\begin{align}
\begin{aligned}
X_1 &\overset{\epsilon}{\rightarrow} X_2\\
X_2 &\overset{\epsilon}{\rightarrow} X_1
\end{aligned}
\end{align}
is the spontaneous switching of food source by an ant.
\smallskip

\begin{figure}
 \includegraphics[width=0.95\linewidth]{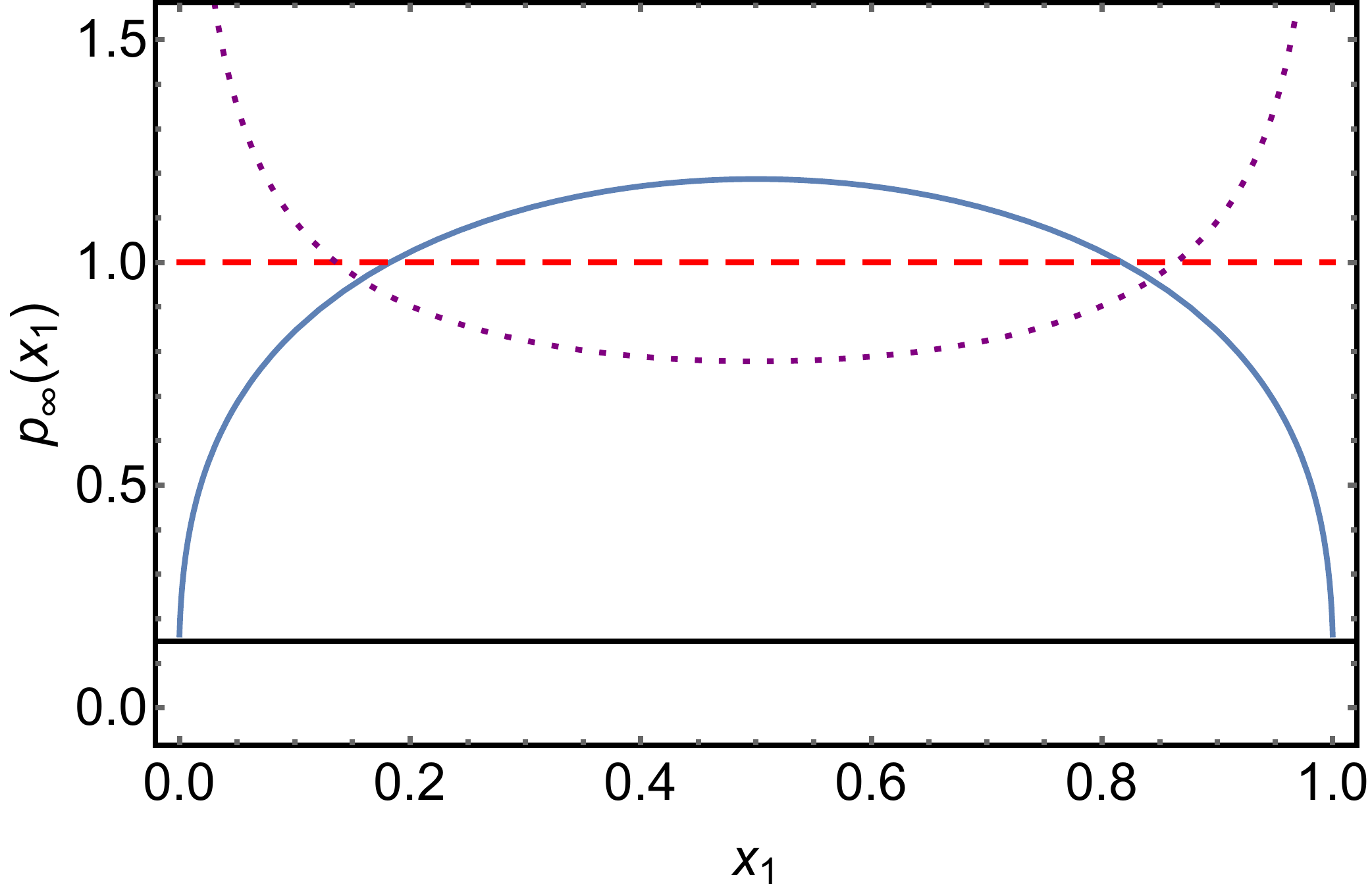}
\caption{Solutions of the stationary FPE for the foraging colony model for different parameter values. We used $N=1000$ and $N_C/N = 0.75$ (blue, solid); $N_C/N = 1$ (red, dashed); $N_C/N = 1.5$ (purple, dotted). Depending on the parameters, the system exhibits either a bistable or a monostable behavior.}
\label{fig:ants:fpe}
\end{figure}

The authors in \cite{mcKane_ameisen} find that the system exhibits either bistable behavior, where most ants visit the same food source and switch unregularly and within a short time to the other source or a monostable behavior where both food sources are frequented by approximately half the ants. Whether mono- or bistable behavior occurs depends on the quantity
\begin{equation}
N_C = \frac{r}{\epsilon} \; ,
\label{eq:n_c}
\end{equation}
with $N_C \gg N$ implying bistability and $N_C \ll N$ monostability. In particular, monostable behavior is the only type of behavior found in the  deterministic limit of very large colony sizes $N$. 
Since the total number of ants is constant, the number of ants of species 2 is given by $n_2=N-n_1$. This is equivalent to the condition for the species concentrations $x_1 + x_2 = 1$. As a result of this conservation law, the FPE of this system can be reduced to one-dimension. The authors solved it analytically in the stationary case, which leads to the probability distributions shown in Fig.~\ref{fig:ants:fpe}. 

\smallskip

Instead of solving the FPE explicitly, we will calculate the favorable states directly using (\ref{eq:alpha}) and show the SPP of the system. Even though the model is effectively one-dimensional, we will always use both variables, $x_1$ and $x_2$. 
\smallskip

Drift and diffusion are given by
\begin{align}
\vec{f}(x_1, x_2) &= 
\begin{pmatrix}
\epsilon(x_2-x_1)\\
-\epsilon(x_2-x_1)\\
\end{pmatrix}
\label{eq:ants:drift}
\end{align}
and
\begin{align}
\mathbf{D} &= 
\begin{pmatrix}
2 r x_1 x_2 + \epsilon & -2 r x_1 x_2 - \epsilon\\
-2 r x_1 x_2 - \epsilon & 2 r x_1 x_2 + \epsilon
\end{pmatrix}\, .
\end{align}

This leads to a convection flow of
\begin{align}
\vec{\alpha}(x_1,x_2) = 
\begin{pmatrix}
\left(\epsilon - \frac{r}{N}\right)(x_2-x_1)\\
-\left(\epsilon - \frac{r}{N}\right)(x_2-x_1)\\
\end{pmatrix}\, .
\label{eq:ants:alpha}
\end{align}

The roots of (\ref{eq:ants:alpha}) are $x_1 = x_2 = 1/2$. 
This means that we only expect one extremum of the stationary probability distribution, which lies at the same position as the deterministic fixed point that can be calculated from \eqref{eq:ants:drift}. However, this does not mean that the stochastic behavior of the system is similar to the deterministic one: The Jacobian of the deterministic system (\ref{eq:ants:drift}) is
\begin{equation}
\mathbf{J_{\text{det}}} = 
\begin{pmatrix}
-\epsilon & \epsilon\\
\epsilon & -\epsilon
\end{pmatrix}\; 
\end{equation}
with eigenvalues $\lambda \in \{0, -2\epsilon\}$. The deterministic attractor of the system is therefore a stable fixed point line at $x_1 = x_2 = 1/2$.

The stochastic Jacobian, which is based on $\vec \alpha$, reads though
\begin{equation}
\mathbf{J_{s}} = 
\begin{pmatrix}
-\left(\epsilon - \frac{r}{N}\right) & \left(\epsilon - \frac{r}{N}\right)\\
\left(\epsilon - \frac{r}{N}\right) & -\left(\epsilon - \frac{r}{N}\right)
\end{pmatrix}\; ,
\label{eq:ants:js}
\end{equation}
with eigenvalues $\lambda \in \{0, -2(\epsilon-r/N)\}$. The stochastic fixed point at $x_1 = x_2 = 1/2$ is only stable for $r/N < \epsilon$. Otherwise, the fixed point  is a minimum, and the maxima of the probability distribution are located at the boundary. This is exactly the same result as obtained from the full solution of the FPE in \eqref{eq:n_c}. The fact that favorable states (or maxima of the probability distribution) can also emerge at the boundary of the state space, even though the deterministic drift at this points is not zero, can be interpreted as a special feature of stochastic systems which cannot occur in a deterministic description. We will discuss this in more detail in our second example. 
% \tdBD{das stimmt doch nicht: es gibt doch in Reaktionssystemen und Populationsdynamik die Randfixpunkte, bei denen eine oder mehr Spezies ausgestorben sind.}

Of course, our findings so far provide no new insights into the ant model compared to the original investigation in \cite{mcKane_ameisen}. We want to note though that our results can be obtained without the need to solve any differential equation and by following a formalism that is very similar to the deterministic analysis of the model.
\smallskip

\begin{figure}
    \subfigure[$r = 10, \epsilon = 100$]{\includegraphics[width=0.49\linewidth]{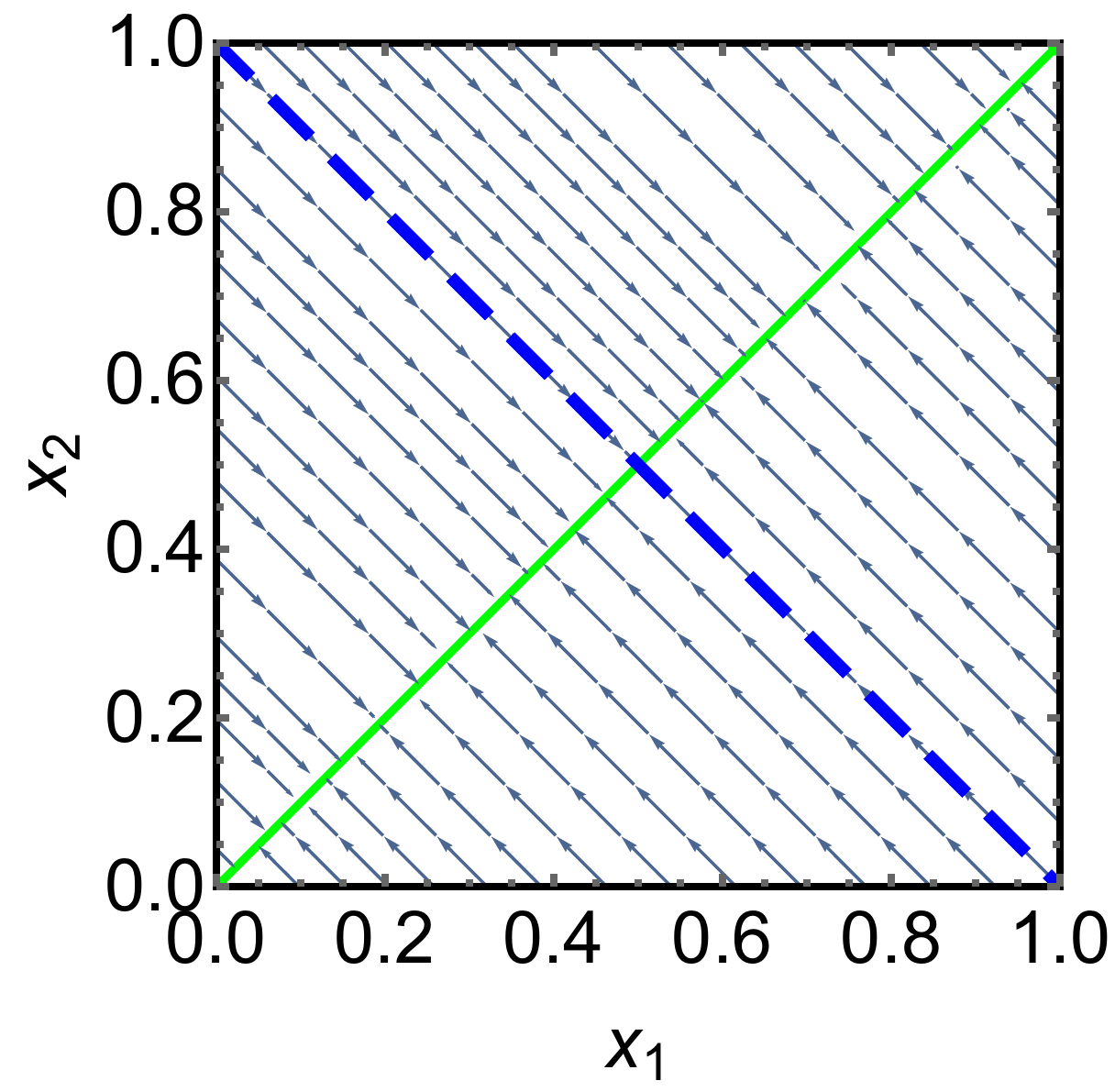}}
    \subfigure[$r = 100, \epsilon = 10$]{\includegraphics[width=0.49\linewidth]{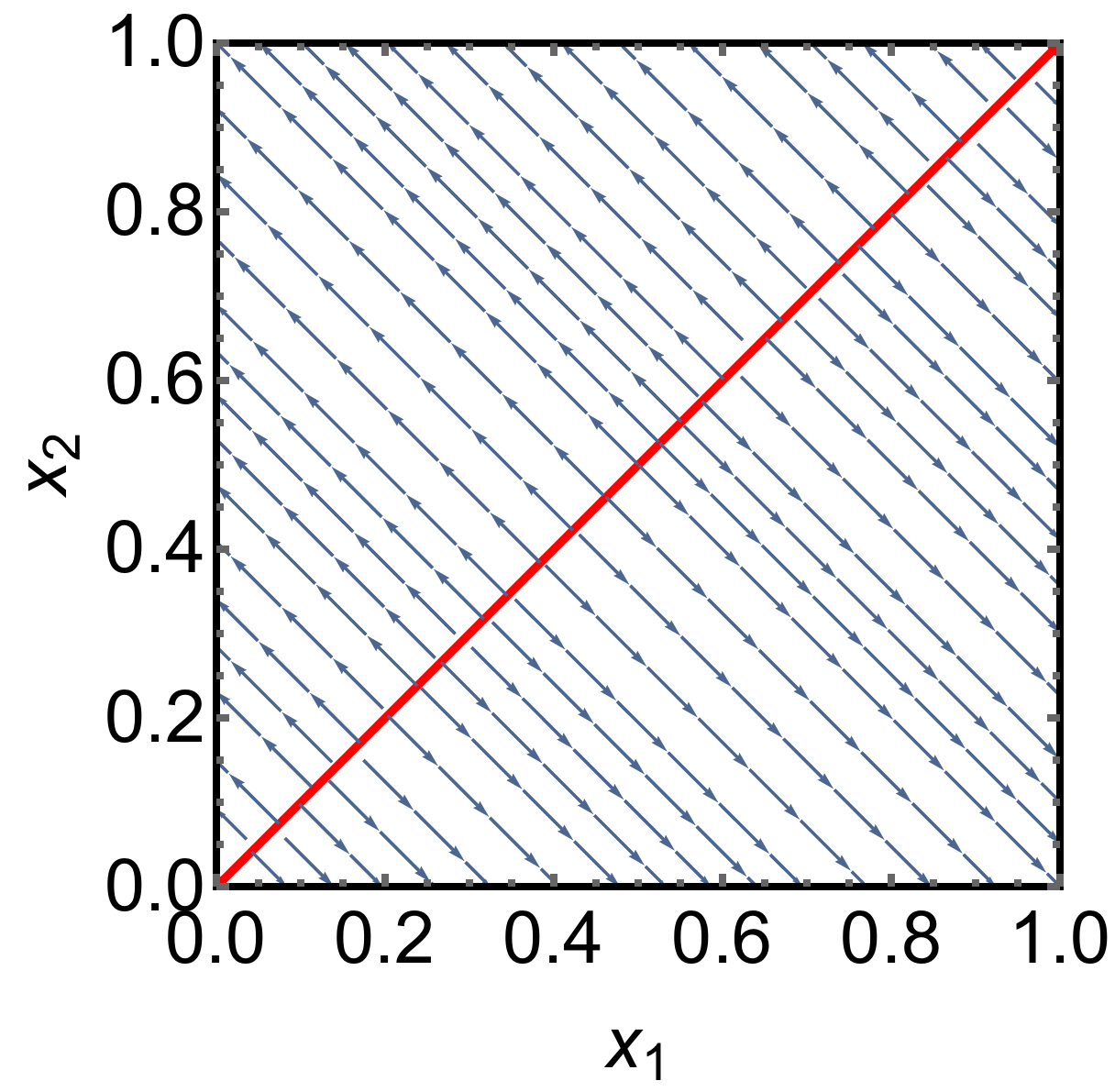}}
\caption{SPPs for the ant model for different parameter values. For $r/N < \epsilon$ (left) we obtain a stable fixed point line just as in the deterministic system. For $r/N > \epsilon$ however, the fixed point line becomes unstable,  and the points of maximum probability are at the boundary of the state space. Due to the conserved total number of ants, the dynamic of both systems is constrained to a line with $x_1 + x_2 = 1$, which is implied by the dashed blue line in the left figure.}
\label{fig:ants:phase_space}
\end{figure}

Based on \eqref{eq:ants:alpha} and the stability of the fixed point from \eqref{eq:ants:js}, we are now able to draw the SPP for the ant model, which is depicted in Fig.~\ref{fig:ants:phase_space}.

\subsection{Example: Rosenzweig-MacArthur model}
\label{sec:rosenzweig}
\subsubsection{General properties}
Next, we focus on a truly two-dimensional model that can show limit cycles in the deterministic version. This is the so-called Rosenzweig-MacArthur model for predator-prey interactions \cite{rosenzweig}. It is usually given in terms of ODEs, but it can be also represented by the following set of reactions:

\begin{align}
\begin{aligned}
X & \overset{\beta}{\longrightarrow} 2 X &\text{prey birth}\\
X + X &\overset{\delta}{\longrightarrow} X & \text{prey competition}\\
X + Y &\overset{\frac{d}{1+Ax}}{\longrightarrow} 2 Y & \text{predator-prey interaction}\\
X + Y &\overset{\frac{b-d}{1+Ax}}{\longrightarrow} Y & \text{predator-prey interaction with loss}\\
Y &\overset{c}{\longrightarrow} \emptyset & \text{predator death}\\
\emptyset &\overset{q}{\longrightarrow} X & \text{prey immigration}\\
\emptyset &\overset{q}{\longrightarrow} Y & \text{predator immigration}
\label{eq:rosenzweig:reactions}
\end{aligned}
\end{align}

The immigration reactions are only introduced to prevent spontaneous extinctions (the so-called Keizer's paradox \cite{keizer}) and can be neglected for the deterministic analysis. Drift vector and diffusion matrix for this system are easy to calculate and read
\begin{equation}
\vec{f} = 
\begin{pmatrix}
q - R \cdot \frac{b}{d} + x \left(\beta - \delta (x-\frac{1}{N})\right) \\
q +R - c y
\label{eq:rosenzweig:drift}
\end{pmatrix}
\end{equation}
and
\begin{equation*}
\mathbf{D} = 
\begin{pmatrix}
q + R \cdot \frac{b}{d} + x(x-\frac{1}{N})\delta + x \beta & -R\\
-R & q + R + c y
\end{pmatrix}
\end{equation*}
with the abbreviation $R = \frac{d x y}{1 + A x}$.
\smallskip

% After some rescaling that is described in \cite{bazykin}, we can  rewrite the deterministic dynamics given by \eqref{eq:rosenzweig:drift} in the following form:
% \begin{align}
% \begin{aligned}
% \dot{u}&= u-\frac{u v}{1 + u \kappa}-u^2 \epsilon& \text{Prey} \phantom{rirr}\\
% \dot{v}&=-v \gamma \left(1-\frac{u}{1+u\kappa}\right)& \text{Predator}
% \label{eq:multidimensional:hopf:umskaliert}
% \end{aligned}
% \end{align}

% With rescaled time and populations $\tau = \beta t, u = \frac{d}{c}x, v = \frac{b}{\beta}y$ and rescaled parameters $\kappa = Ac/d, \gamma = c/\beta$ and $\epsilon = \frac{\delta c}{\beta d}$.
% \tdBD{die mehrfachen Notationswechsel sind extrem verwirrend. Man kann doch bei den originalvariablen (Individuendichten x und y) bleiben und trotzdem argumentieren, dass die Bifurkationen nur von effektiven Parametern kappa und epsilon abhaengen. Man kann auch gleich oben im Reaktionssystem beta=1 waehlen und sagen, dass man die Zeitskala durch die Rate der ersten Reaktion festsetzt. ICh finde, dass auf jeden Fall das normale Rosenzweig-McArthurmodell in den Variablen x und y hier stehen sollte statt in den Variablen u und v. Dann kann man auch schoen das f aus (45) ablesen. }

\begin{figure}
 \subfigure[(Deterministic) phase portrait for the Rosenzweig-MacArthur model. Blue arrows indicate the drift $\vec{f}(\vec{x})$. The black lines are the nullclines where one component of $\vec f$ vanishes. Parameters are $A = 2/3; d = 0.65; c = 0.65; b=1; q=0.01; \delta = 0.2; \beta = 0.8$. The fixed point (green dot) is a stable spiral. The red line shows a sample trajectory.]{\includegraphics[width=0.85\linewidth]{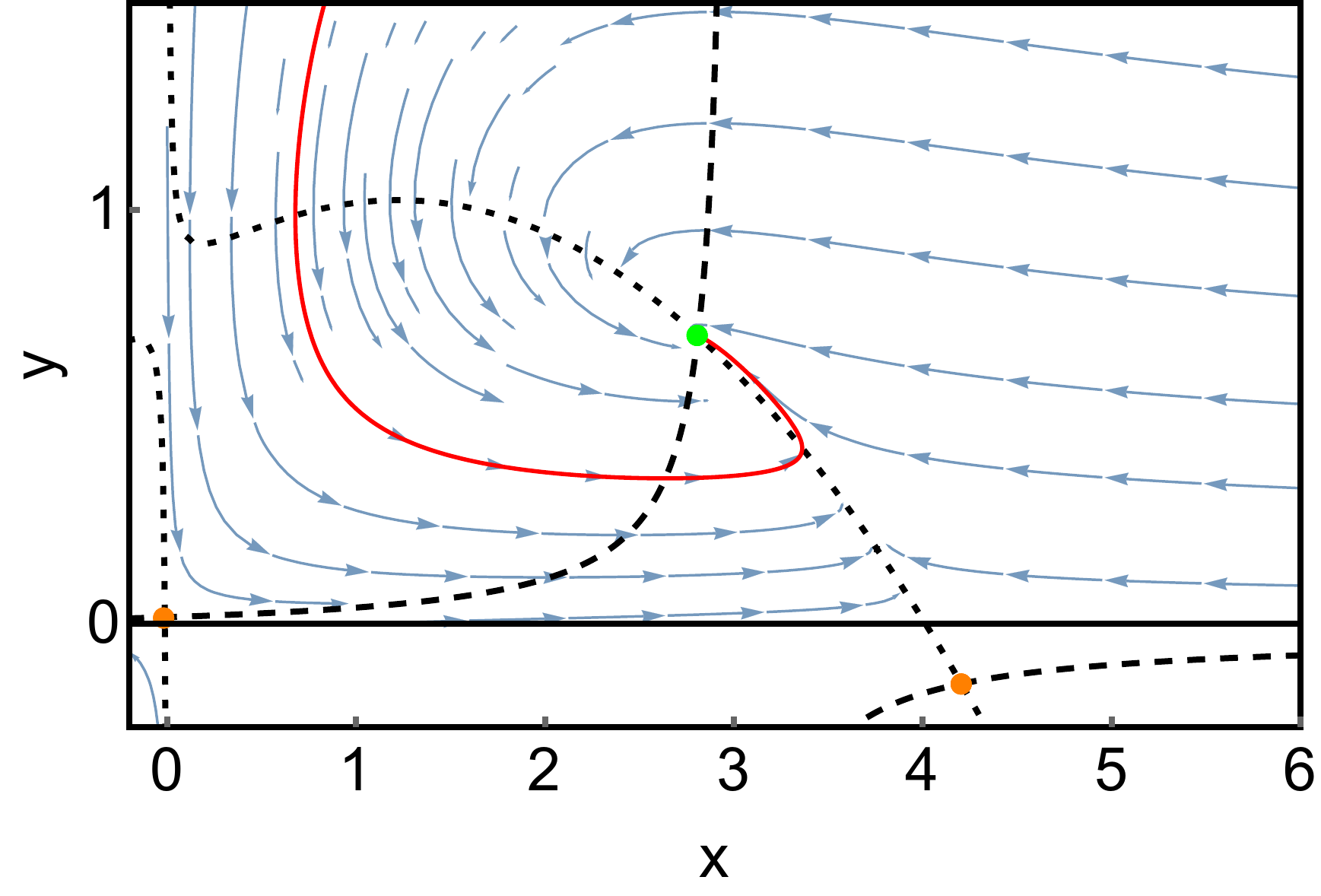}}
  \subfigure[Parameters from (a) are changed to $\delta = 0.1; \beta = 0.9$.  The system exhibits now an unstable spiral (red dot) enclosed by a limit cycle, to which the shown red sample trajectory converges.]{\includegraphics[width=0.85\linewidth]{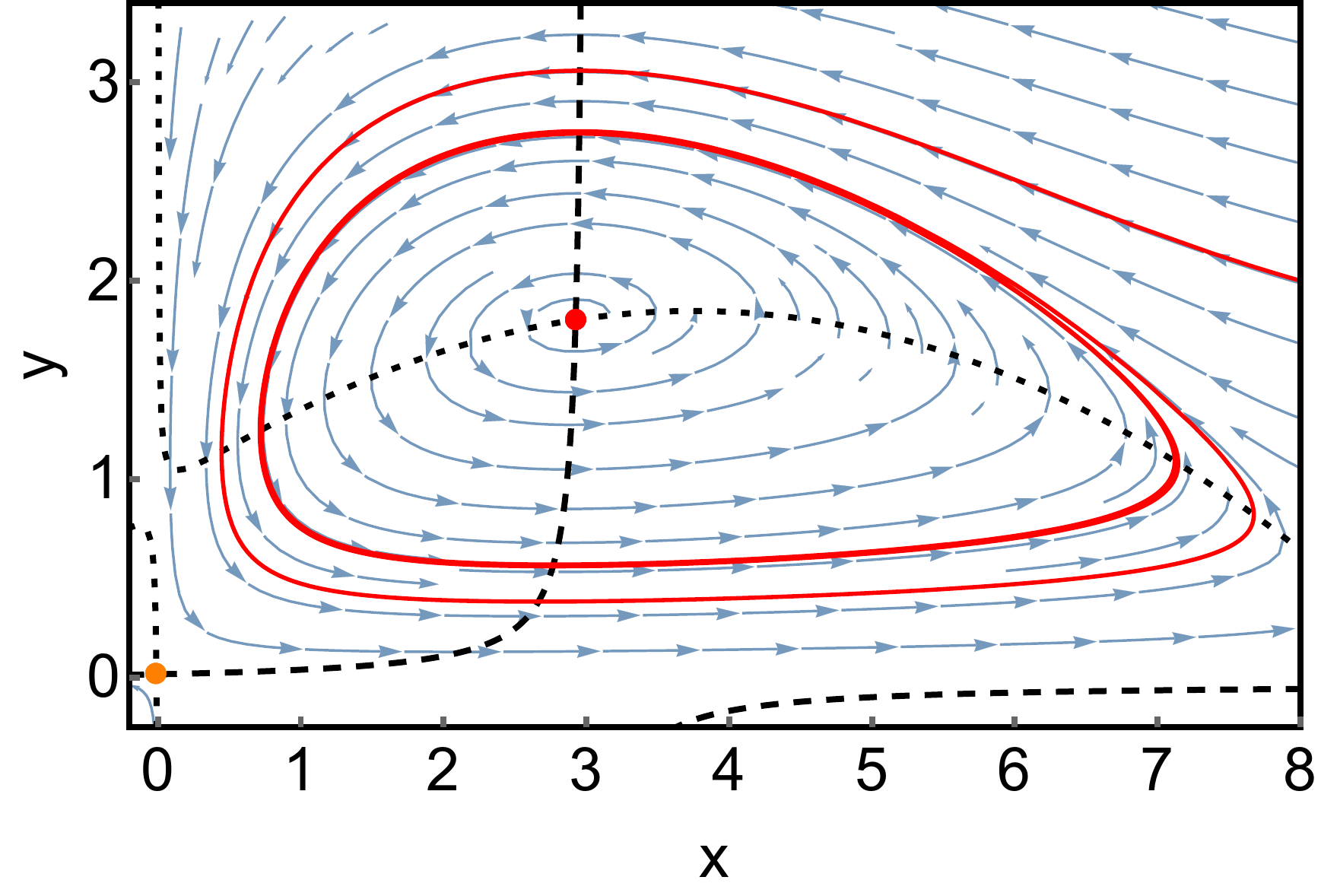}}
   \subfigure[Stability diagram of the rescaled Rosenzweig-MacArthur model.]{\includegraphics[width=0.85\linewidth]{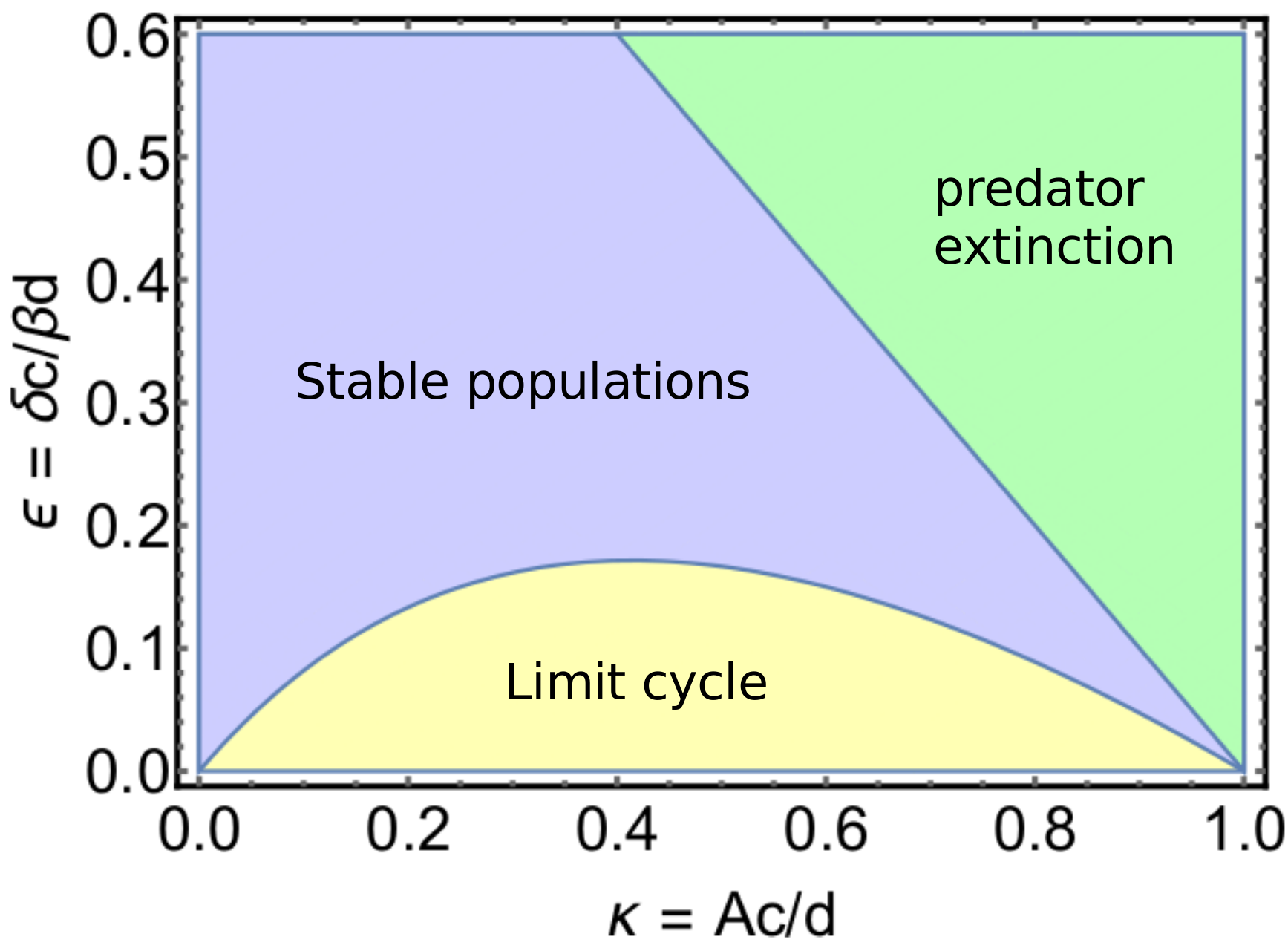}}
\caption{}
\label{fig:rosenzweig:stability}
\end{figure}

Depending on the effective parameters  $\kappa = Ac/d$ and $\epsilon = \frac{\delta c}{\beta d}$, the deterministic system corresponding to \eqref{eq:rosenzweig:drift} can either exhibit stable populations of both predator and prey, a limit cycle where the populations of both species oscillate, or the extinction of the predator\cite{Schmitt2014, bazykin}. A phase diagram showing the parameter dependency of these regimes is depicted in Fig.~\ref{fig:rosenzweig:stability} (c), along with the (deterministic) phase portraits for two sets of parameter values in (a) and (b).

\subsubsection{The effect of stochasticity when the deterministic system has a stable fixed point}

\begin{figure}
 \subfigure[$N = 1000$]{\includegraphics[width=0.52\linewidth]{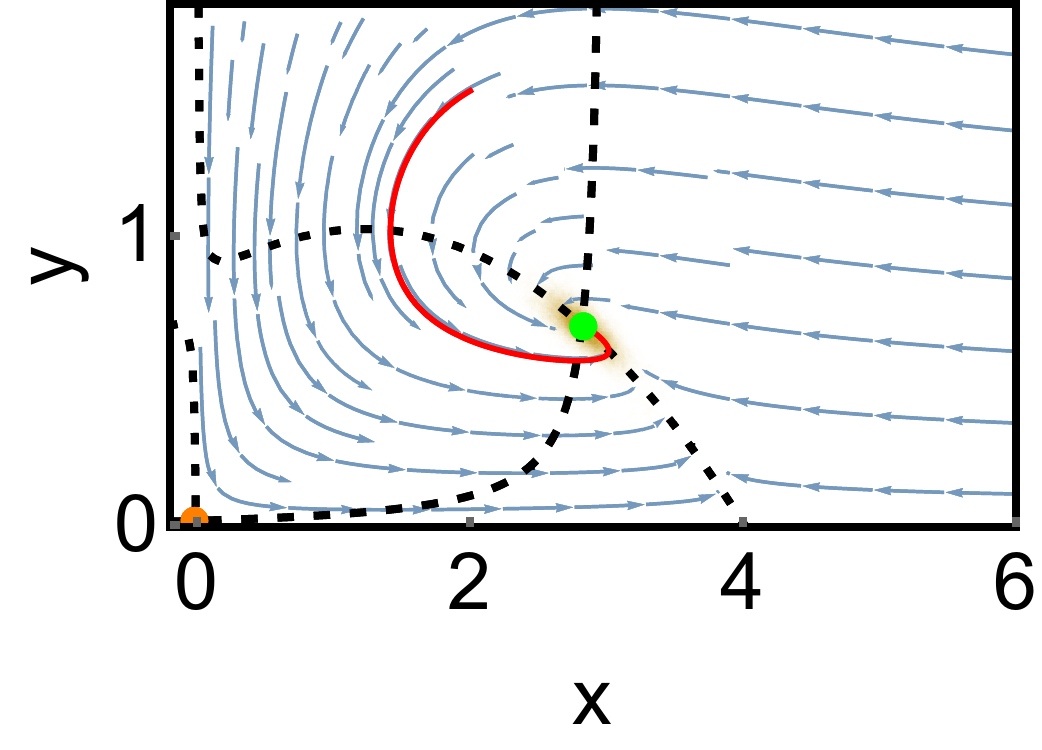}}
  \subfigure[$N = 80$]{\includegraphics[width=0.48\linewidth]{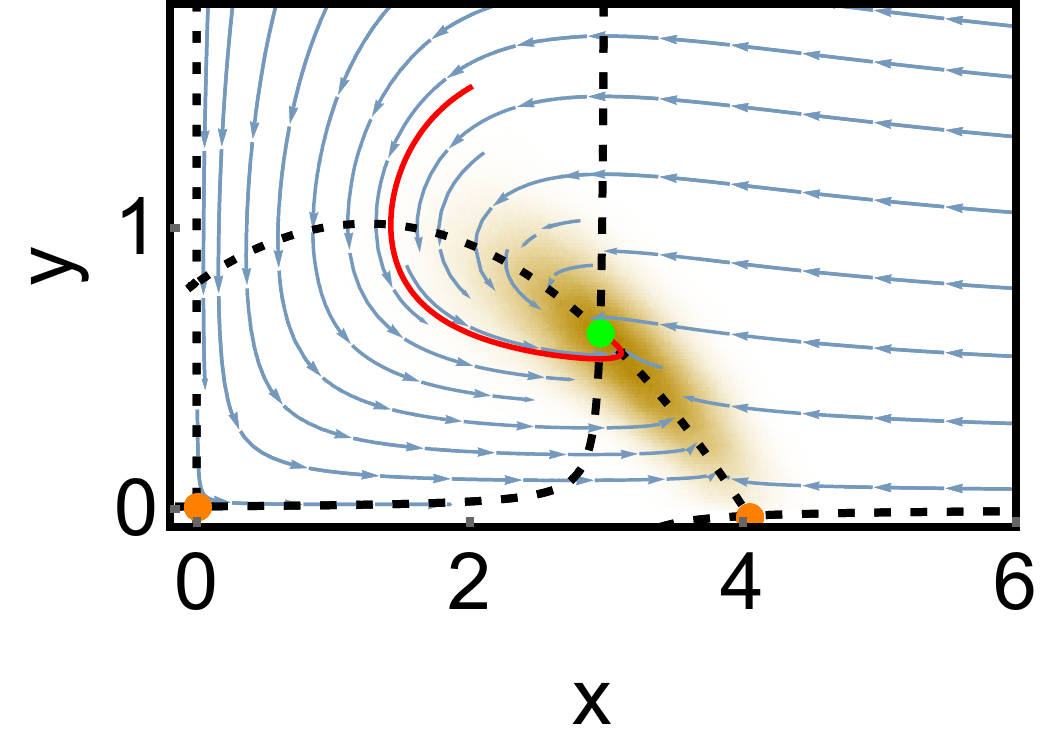}}
 \subfigure[$N = 50$]{\includegraphics[width=0.48\linewidth]{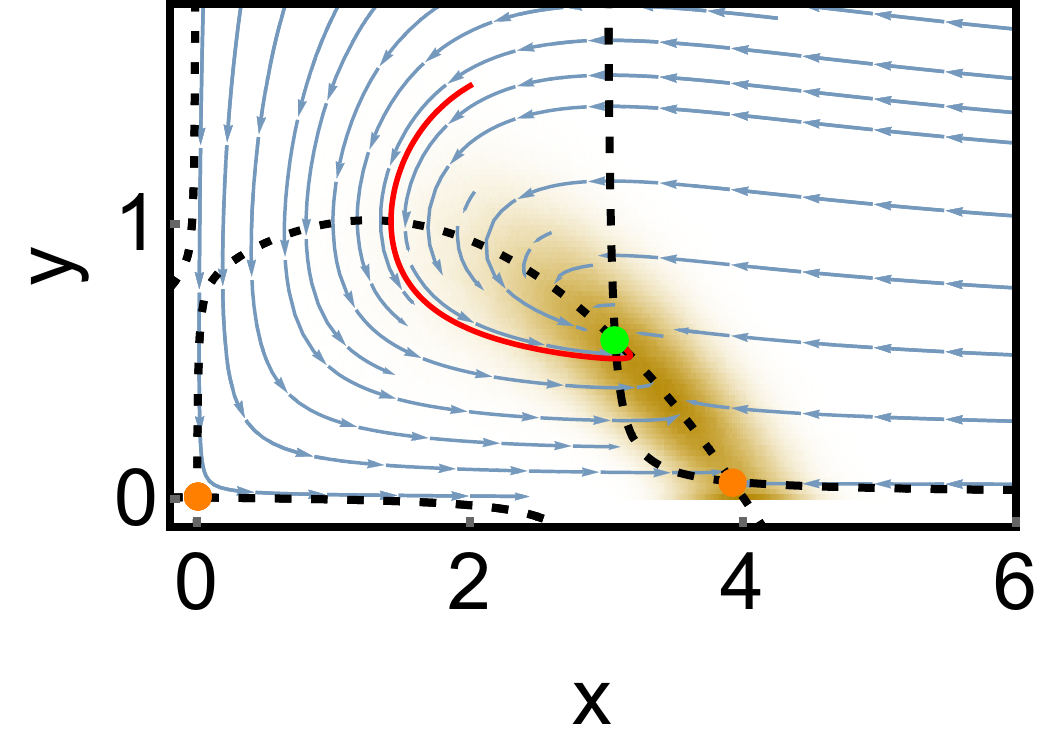}}
 \subfigure[$N = 35$]{\includegraphics[width=0.48\linewidth]{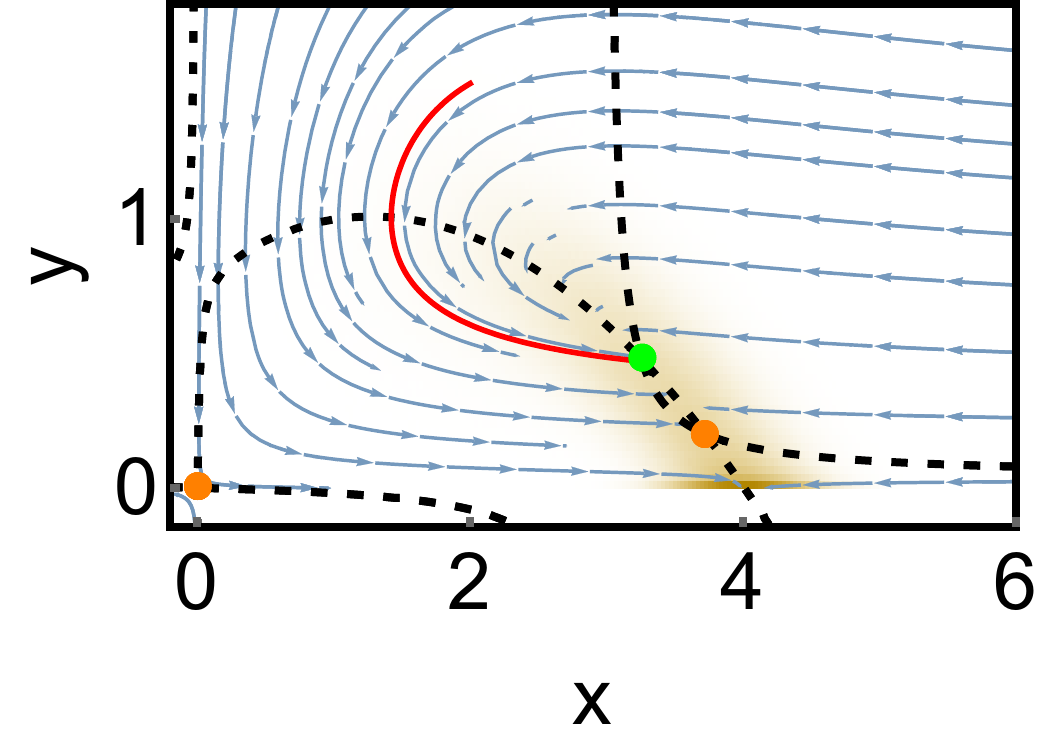}}
  \subfigure[$N = 10$]{\includegraphics[width=0.48\linewidth]{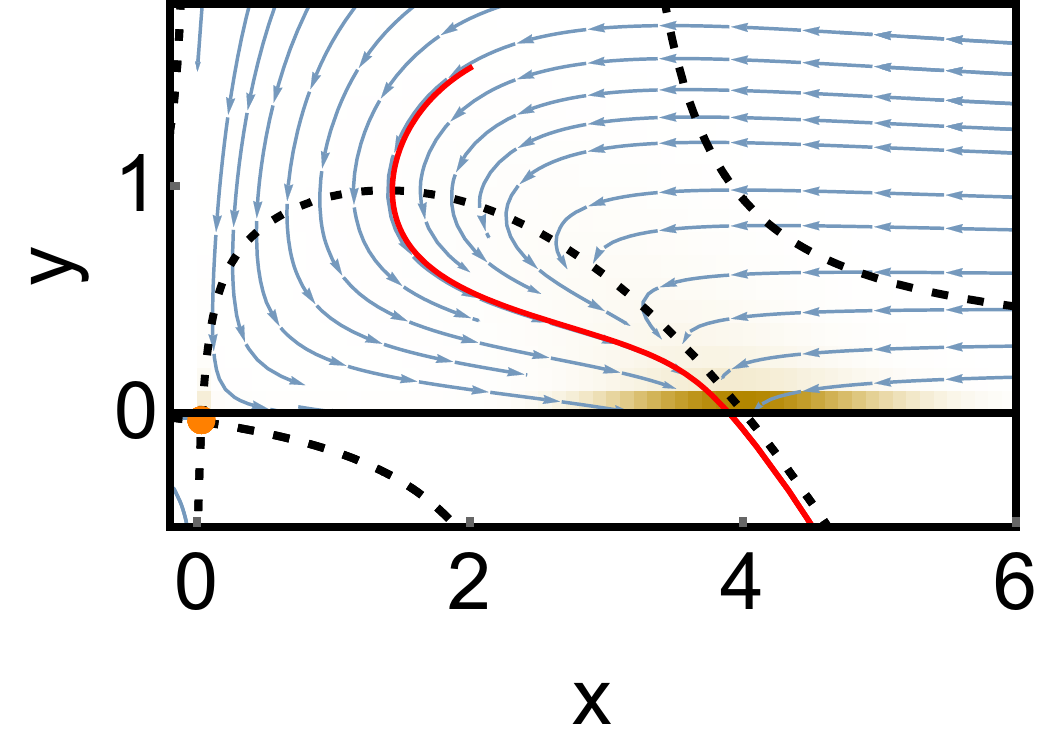}}
\caption{SPPs of the Rosenzweig-McArthur model for the case that the deterministic system has a stable fixed point,  for different system sizes. Parameters are chosen as in Fig.~\ref{fig:rosenzweig:stability}(a). The yellow colour shade indicates the density of points  obtained from a stochastic simulation of the same model. The other colours are the same as in Fig.~\ref{fig:rosenzweig:stability}.}
\label{fig:rosenzweig:stabil:phasenraum}
\end{figure}

Figure~\ref{fig:rosenzweig:stabil:phasenraum} shows SPPs for different system sizes $N$ with reaction rates chosen such that the deterministic system has a stable fixed point.  In order to assess the significance of the SPPs, we also performed Gillespie simulations of the corresponding reaction system using the free software tool \textit{Dizzy Gillespie} \cite{dizzy}. The time series data from these simulations was used to generate density histograms, which correspond to the respective stationary probability distributions. These  are shown as yellow clouds overlayed over the SPPs.

\medskip

For the large system size $N=1000$, the SPP shown in  Fig.~\ref{fig:rosenzweig:stabil:phasenraum}(a) agrees with the deterministic phase portrait in Fig.~\ref{fig:rosenzweig:stability}(a). Fluctuations around the maximum value, which are obtained from the stochastic simulation, are rather small. For the smaller system size of $N = 80$, we obtain a similar SPP, however, the probability distribution obtained from the stochastic simulation is broadened due to the increasing noise. The maximum of the probability distribution coincides well with the stable fixed point of the SPP. For $N = 50$, the SPP remains unchanged, but the stochastic data become qualitatively different:  While there is only one density maximum at approximately $(3, 0.7)$ for $N = 80$, there emerges a second maximum at the boundary of the phase space at approximately $(4, 0)$ for $N=50$. We can interpret this emergence of a second maximum as a stochastic bifurcation, which we will discuss in more detail in the next subsection. For $N = 35$, the boundary maximum becomes more important, while the central maximum is almost gone. At $N = 10$, a second bifurcation has occurred, which is directly visible in the SPP: As the two nullclines cease to intersect, the central maximum vanishes and all probability becomes concentrated close to the boundary maximum at $x = 4$. 
\smallskip

The SPP for $N = 10$ provides also a very clear understanding of how the maximum at the boundary emerges: As no fixed point is left inside the phase space the probability follows the convective field until a boundary (in this case the x-axis) is reached. From this position, the probability flow can't follow the convective field anymore since negative values must retain probability zero. Therefore, the probability flow slides along the x-axis until the x-component of the convective field equals zero. Naturally, this happens at the position where the x-nullcline intersects the x-axis. The boundary maxima emerge therefore always at the intersection points of a nullcline with its corresponding axis.

\subsubsection{Nullcline gap bifurcation}
\label{sec:nullcline_gap}
\begin{figure}
 \subfigure[$N \gg N_C$]{\includegraphics[width=0.49\linewidth]{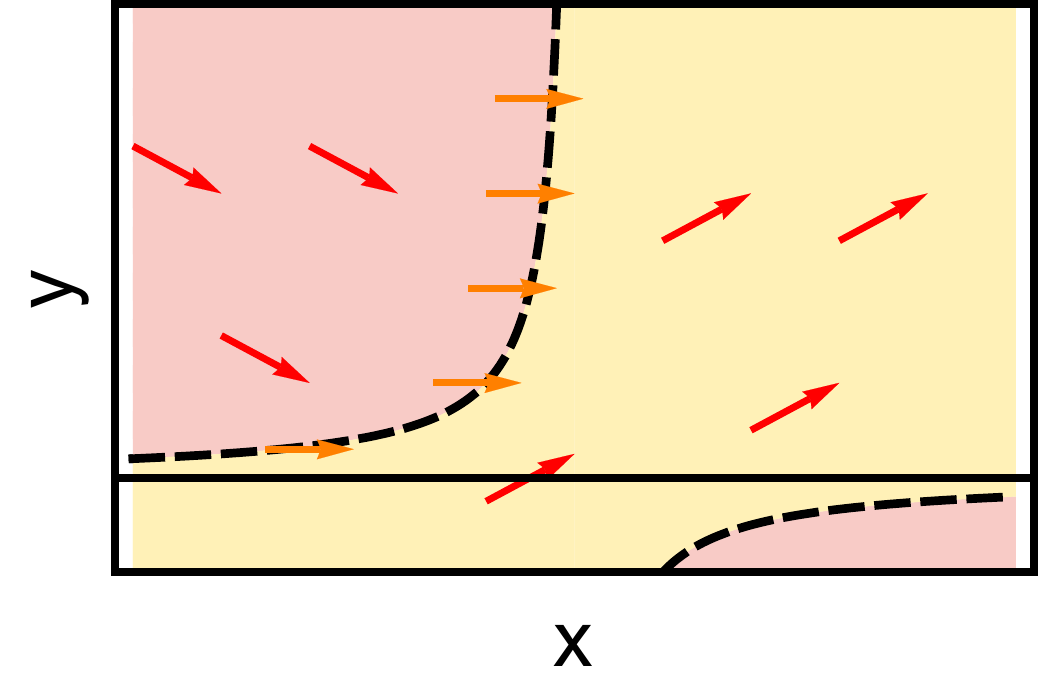}}
 \subfigure[$N \ll N_C $]{\includegraphics[width=0.49\linewidth]{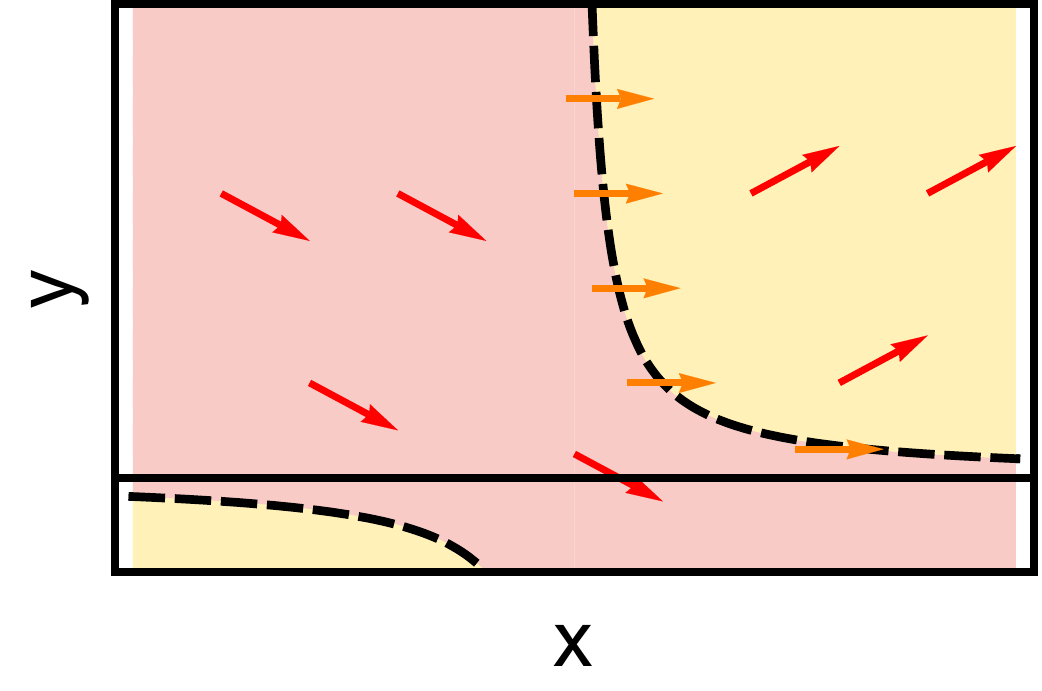}}
 
\caption{Schematic SPPs illustrating the nullcline gap bifurcation. The nullclines (black, dashed) determine the positions in state space where the arrows of the convective field are completely horizontal (orange arrows). Thereby they divide the state space into two distinct regions where all those arrows point either upwards (yellow region) or downwards (red region). Depending on the region in which the gap occurs probability can either escape through the gap and reach the boundary of the state space ($N \ll N_C$, Fig.~(b)) or is caught inside ($N \gg N_C$, Fig.~(a)) so that no boundary maxima can emerge.
}
\label{fig:rosenzweig:nullcline_gap}
\end{figure}

With this understanding of the boundary maxima, we reconsider the bifurcation that leads to their formation, which occurs for our system between $N = 80$ and $N = 50$. Comparing the SPPs for those two system sizes (Fig.~\ref{fig:rosenzweig:stabil:phasenraum}(b) and (c)), one sees that the two predator nullclines at the bottom exchange their direction.
This is also illustrated in the exaggerated scheme in Fig.~\ref{fig:rosenzweig:nullcline_gap}. The dashed lines indicate the y-nullclines. As their behavior close to the x-axis changes qualitatively, the direction of the convective flow also changes: When the convection vectors in the gap point upwards (Fig.~\ref{fig:rosenzweig:nullcline_gap}a), the probability moves away from the $x$-axis and cannot produce a boundary maximum.  When the vectors point downwards (Fig.~\ref{fig:rosenzweig:nullcline_gap}b), the probability flow in the gap region moves towards the axis, and a boundary maximum will emerge. 
\smallskip

Since the formation of a boundary maximum is directly coupled to the change of the nullcline gap, we call this type of bifurcation a \textit{nullcline gap bifurcation}.

\subsubsection{The effect of stochasticity when the deterministic system has a limit cycle}
\begin{figure}
 \subfigure[$N = 1000$]{\includegraphics[width=0.49\linewidth]{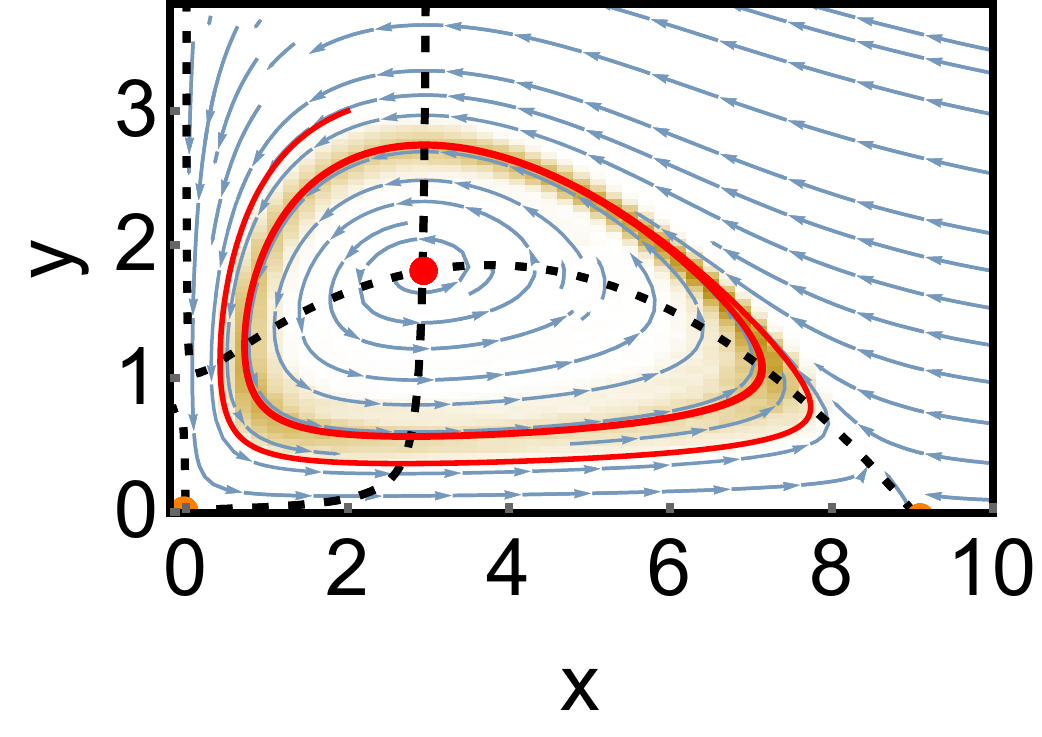}}
 \subfigure[$N = 80$]{\includegraphics[width=0.49\linewidth]{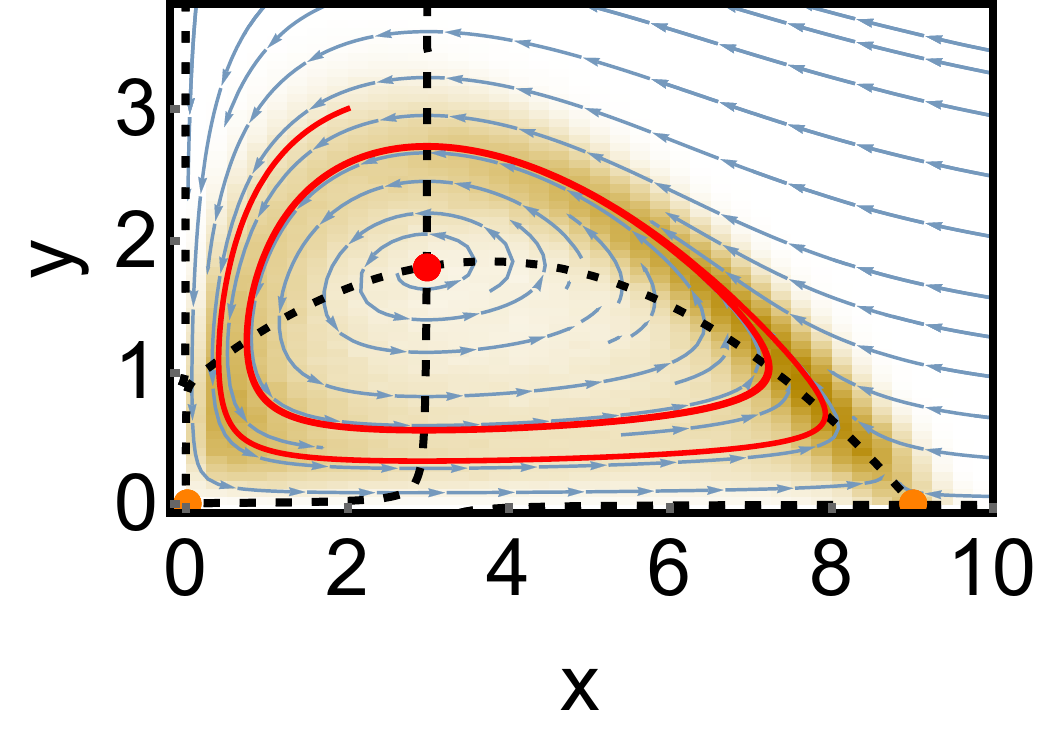}}
 \subfigure[$N = 50$]{\includegraphics[width=0.49\linewidth]{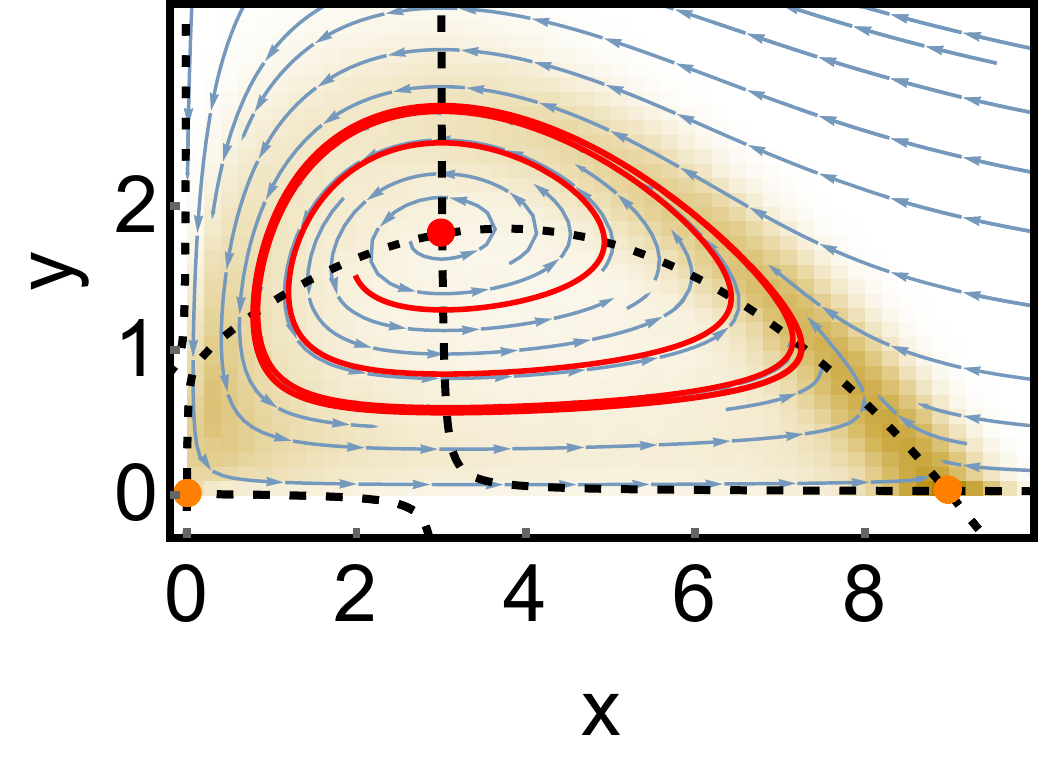}} 
\caption{SPPs of the Rosenzweig-McArthur model for the case that the deterministic system has a stable fixed point,  for different system sizes. Parameters are chosen as in Fig.~\ref{fig:rosenzweig:stability}(b). The colours are the same as in the previous figures. The yellow shade indicates again the stationary probability distribution obtained from a stochastic simulation.}
\label{fig:rosenzweig:instabil:phasenraum}
\end{figure}

Figure~\ref{fig:rosenzweig:instabil:phasenraum} shows SPPs for the parameter set that yields a limit cycle in the deterministic model.

Again, there is a good agreement between the SPP shown in Fig.~\ref{fig:rosenzweig:instabil:phasenraum}(a) and the deterministic phase portrait shown in Fig.~\ref{fig:rosenzweig:stability}(b). The stationary probability obtained from the stochastic simulation is concentrated at the location of the limit cycle of the SPP. A closer look reveals though that some states have a much higher probability than others (indicated by a darker yellow coloring). These are the slow transient states discussed in section \ref{sec:favorable_states}, where $p_{\infty}$ has maxima with non-vanishing probability current. 

\smallskip
For $N = 80$, the probability distribution again becomes broader. Around $N = 65$  the Nullcline Gap bifurcation occurs, leading to a boundary maximum at $x \approx 9$ for the smaller system size of $N=50$.

% \subsubsection{Investigating the Hopf bifurcation}
% \tdBD{dieser Abschnitt muss entweder durch Daten belegt werden, oder gestrichen werden. ICh bin fuer Streichen. }
% One would now naturally investigate the influences of the intrinsic noise on the Hopf bifurcation which is the transition between both parameter sets that we investigated so far. Hereby the limit cycle gets smaller and smaller until it merges with the instable spiral and leaves the stable spiral behind. It proves however rather difficult to do so: On one hand, it is not easy to obtain the exact parameter values at which the bifurcation happens from simulations as both a stable spirale and a small radius limit cycle close the Hopf bifurcation have almost identical probability distributions. Therefore it is only possible to obtain the critical value of the bifurcation parameter with very low accuracy from simulations. On the other hand this critical value deviates only very little from the deterministic one if the intrinsic noise is small, i.e. the system size is rather big. However, if $N$ is decreased too much, the nullcline gap bifurcation happens inevitably, letting most of the probability escape towards a boundary maximum. Hereby it becomes even more difficult to analyse the little remaining probability at the bifurcation site and thus compare the predictions of the SPPs with stochastic simulations. We therefore came to the conclusion that our version of the Rosenzweig-McArthur model is not suitable to investigate the effects of intrinsic noise on the Hopf bifurcation.

\section{Discussion}    
We have presented a method to obtain phase portraits for stochastic systems (SPPs) as vector plots of the convective field obtained from the Fokker-Planck equation (FPE). This field can be obtained directly from the drift and diffusion terms without the need to solve the FPE. We showed that stable (unstable) fixed points of the convective field correspond to maxima (minima) of the stationary probability distribution if the probability flow vanishes at this point. This means that for a one-dimensional system the SPPs always reproduce correctly the maxima and minima of the stationary probability distribution (as obtained from the Fokker-Planck equation). We demonstrated this using a model of foraging ants \cite{mcKane_ameisen}, for which the SPP shows the change of stability of the symmetric fixed point as the system size is changed. Earlier, an explicit solution of the FPE was required in order to derive this result.    

Furthermore, we considered a predator-prey model \cite{rosenzweig}, for which the stationary probability flow does not vanish. We could also verify that the SPPs yield an accurate description of the stochastic dynamics of the system, which were obtained from Gillespie simulations. In particular, we found that boundary maxima of the stationary probability distribution can also be predicted from the SPPs. The emergence of these boundary maxima happens as a special type of bifurcation which can only appear in stochastic systems, and which we called nullcline-gap bifurcation.

These successes of SPPs suggest that they are for many systems a good indicator of where the probability concentrates in the stationary state. However,  we were so far not able to determine whether a fixed point of the convective field $\vec{\alpha}$ always leads to an extremum in the stationary probability distribution for multidimensional systems. So far, we did not yet find a counterexample. Similarly, it is unclear whether limit cycles in the SPPs  correspond always to a non-equilibrium steady-state with $\vec{j}\ne 0$ of the underlying stochastic system. Furthermore,
although our example showed a higher concentration of the probability density along the limit cycle, we could not prove that limit cycles in the SPP are always associated  with a crater-shaped stationary probability distribution, and if so, whether the ridge of this crater will always coincide with the course of a limit cycle in the SPP. In any case, as the SPPs are based on the Fokker-Planck equation and not on the Master equation, the agreement of the SPPs with the results of stochastic simulations cannot be better than that of the stationary distribution resulting from Fokker-Planck equation with the true stationary distribution. 

%Taking all this unknowns into account it seems kind of surprising that the SPPs work so well in predicting the favorable states and periodic solutions of the Fokker-Planck-Equation. It would therefore be very interesting to develop a deeper understanding of the underlying principles. In this context, it would also be helpful to search for examples where the SPPs fail to predict the favorable states of the underlying stochastic system correctly.

The results obtained so far suggest further directions of study. In particular, it would be helpful if one could extract information about the relative height of the maxima of the stationary probability distribution from the SPPs.  We have seen examples where the SPP suggests a maximum at the boundary and one inside the state space, while in reality the boundary maximum was so high that it was almost impossible to notice the small maximum inside the state space. The size of the maxima might be correlated with the relative sizes of the basins of attraction or with the density of the probability flow of the convective field. Similarly, the SPPs give so far no indication of the widths of the respective maxima of the system. We suspect that progress in this direction can be obtained by exploiting the properties of the diffusion matrix.

% \tdBD{hm - es scheint mit aber, dass man aus den SPPs noch mehr Infos holen koennte: die Differenz zwischen $\vec \alpha$ und $\vec f$ koennte doch einen Anhaltspunkt fuer die Breite der Verteilung geben. Und wenn man man aus der Laenge der Pfeile von $\alpha$ die Geschwindigkeit ableitet, hat man auch die Dichteverteilung auf dem Grenzzyklus. Und wenn man schaut, wieviel Fluss in jedes der beiden Maxima geht, weiss man vielleicht auch, welches groesser ist. Man koennte ja mal schauen, wie gross jeweils der Einzugsbereich der Maxima ist. Aber all das wuerde ich nicht mehr in diesem PAper machen. Dieses Paper sollten wir abschliessen und Einreichen. Das Hauptziel hier ist ja, erstmal das Konzept der SPPs einzufuehren.}

Finally, we address the question how our method relates to the  work of Cheng et. al. \cite{most_probable}, where the most probable trajectories of a stochastic system are investigated. The authors  start from a sharp initial probability distribution and calculate analytically the time evolution of the maximum. However, this calculation is based on the premise that the probability distribution shows exactly one maximum at all times $t$. In contrast, our approach has no such restrictions and describes the stationary maxima and minima also precisely. But our approach does not describe transient dynamics, except for the initial dynamics that occur when starting with a flat initial probability distribution. The two approaches are thus complementary to each other. Together, they provide an analytical means to investigate the extrema of multistable Fokker-Planck equations, which are otherwise only solvable by numerical methods.

\label{sec:discussion}

\begin{acknowledgments}
JF acknowledges funding by the Hessen State Ministry of Higher Education, Research and the Arts (HMWK) via the “LOEWE CompuGene” project.

We acknowledge support by the German Research Foundation and the Open Access Publishing Fund of Technische Universität Darmstadt. 
\end{acknowledgments}

\bibliography{references}

\end{document}